\documentclass[journal]{IEEEtran}

\ifCLASSINFOpdf
   \usepackage[pdftex]{graphicx}
   \graphicspath{{../pdf/}{../jpeg/}}
   \DeclareGraphicsExtensions{.pdf,.jpeg,.png}
\else
   \usepackage[dvips]{graphicx}
   \graphicspath{{../eps/}}
   \DeclareGraphicsExtensions{.eps}
\fi

\usepackage{amsmath}
\usepackage{bm}
\usepackage{amssymb}

\begin{document}

\title{A Novel Cross Entropy Approach for Offloading Learning in Mobile Edge Computing}

\author{Shuhan~Zhu,~\IEEEmembership{Student~Member,~IEEE,}
       Wei~Xu,~\IEEEmembership{Senior~Member,~IEEE,}
       Lisheng~Fan,~\IEEEmembership{Member,~IEEE,}
       Kezhi~Wang,~\IEEEmembership{Member,~IEEE,}
      and~George~K.~Karagiannidis,~\IEEEmembership{Fellow,~IEEE}

\thanks{Manuscript received October 1, 2019; revised November 25, 2019; accepted November 28, 2019. This work was supported by the National Key Research and Development Program 2018YFA0701602, by the Natural Science Foundation of Jiangsu Province for Distinguished Young Scholars under Grant BK20190012, by the NSFC under grants 61871109 and 61871139, and by the Royal Academy of Engineering under the Distinguished Visiting Fellowship scheme(DVFS21819$\backslash$9$\backslash$7). \emph{(Corresponding author: Wei Xu.)}}
\thanks{S. Zhu is with the National Mobile Communications Research Laboratory, Southeast University, Nanjing, China (shzhu@seu.edu.cn). }
\thanks{W. Xu is with the National Mobile Communications Research Laboratory, Southeast University, Nanjing, China, and also with Purple Mountain Laboratories, Nanjing, China (wxu@seu.edu.cn).}
\thanks{L. Fan is with the School of Computer Science, Guangzhou University, China (lsfan@gzhu.edu.cn). }
\thanks{K. Wang is with the Department of Computer and Information Sciences, Northumbria University, Newcastle, UK (kezhi.wang@northumbria.ac.uk).}
\thanks{George K. Karagiannidis is with the Aristotle University of Thessaloniki,  Thessaloniki 54 124, Greece (geokarag@auth.gr).}
}

\maketitle

\begin{abstract}
In this paper, we propose a novel offloading learning approach to compromise energy consumption and latency in a multi-tier network with mobile edge computing. In order to solve this integer programming problem, instead of using conventional optimization tools, we apply a \emph{cross entropy} approach with iterative learning of the probability of \emph{elite solution} samples. Compared to existing methods, the proposed one in this network permits a parallel computing architecture and is verified to be computationally very efficient. Specifically, it achieves performance close to the optimal and performs well with different choices of the values of hyperparameters in the proposed learning approach.

\end{abstract}

\IEEEpeerreviewmaketitle

\vspace{-0.7cm}
\section{Introduction}

\IEEEPARstart{W}{ith} the rapid development of electronics and wireless networks, various services are currently supported by modern mobile devices (MD). However, most real-time applications require huge computation efforts from the MDs. Mobile edge computing (MEC) is a very promising technology to solve this dilemma for the next-generation wireless networks. According to MEC, edge servers, which are used to connect mobile terminals with a cloud server, provide high storage capability as well as fast computation ability.

In a MEC architecture both latency and  energy consumption contribute to the network performance and it is of common interest to investigate the problem of balancing these factors with optimized offloading policies. Scanning the open literature, the  authors in \cite{ref1} proposed to offload a task from a single MD to multiple computational access points (CAP). Furthermore, a weighted sum of energy and latency was optimized by using convex optimization. This problem was recently extended in \cite{ref2} to a scenario where multiple MDs perform offloading. The multiple tasks were scheduled based on a queuing state in order to adapt channel variations. Alternatively, in \cite{ref3}, the latency was minimized with scheduling the MEC offloading, while the energy consumption was considered as an individual constraint in MDs.

Recently, machine learning (ML) attracts much attention from both academia and industry, as an efficient tool to solve traditional problems in wireless communication \cite{ref4}-\cite{ref7}. Specifically, the authors in \cite{ref4} proposed a payoff game framework to maximize the network performance through reinforcement learning. Furthermore, a deep Q-network was utilized in \cite{ref5} to optimize the computational offloading, without \emph{a priori} knowledge of the network. Most of these methods, including those which use deep learning network (DNN), focused on the offloading design from a perspective of long-term optimization and at the cost of complexity and robustness \cite{ref6}\cite{ref7}. Moreover, these methods can hardly track fast channel changes, due to the requirement of offline learning. Thus, in general they cannot be applied for real-time applications in time-varying channel and it remains a problem of common interest to optimize offloading policies with a time-efficient method, which simultaneously ensures high-quality performance.

In this work, we introduce the \emph{cross entropy} (CE) approach to solve the offloading association problem, by generating multiple samples and learning the probability distribution of \emph{elite samples}. In contrary to the conventional algorithms, the proposed CE learning approach can use parallel computer architecture to reduce computational complexity, and it works for short-term offloading using online learning architecture, which has a stringent requirement on real-time evaluation. Our work generalizes the CE learning approach to solve the offloading problem with low complexity. The proposed approach is promising since it can effectively replace the traditional convex optimization tools.

\vspace{-0.5cm}
\section{System Model}

We consider the problem of  multi-task offloading in a network with multiple CAPs, where the MD has access to the CAPs. Each of the tasks can be selected to be executed at the local MD or offloaded to the CAPs, while a CAP serves only one task at each time. Since the index `0' represents the local CPU, \({\mathcal N} = \left\{ {1,2, \cdots N} \right\}\) and \({\mathcal M} = \left\{ {0,1, \cdots M} \right\}\) are defined as the sets of tasks and CAPs, respectively. In order to indicate the offloading status, we define the policy
\begin{equation}
{x_{nm}} = \left\{ \begin{array}{l}
1,  \emph{ \emph{if task $n$ offloads to CAP $m$}}\\
0,  \emph{ \emph{otherwise}}
\end{array} \right.
\end{equation}
and matrix \({\bf{X}} = {\left[ {{x_{nm}}|{x_{nm}} \in \left\{ {0,1} \right\}} \right]_{N \times \left( {M + 1} \right)}}\) ensembles all the indices. Assume that each task can be offloaded to a single CPU. In this case holds
\begin{equation}
\sum\limits_{m = 0}^M {{x_{nm}}}  = 1.\label{task allocation}
\end{equation}

\vspace{-0.5cm}
\subsection{Latency}

The execution latency consists of two components: transmission latency and computation time. The transmission time includes task data preparation at the MD, data transmission duration over the air, and received data processing at CAP before conducting computation. Also, the transmission time depends on the achievable rate of physical links. The uplink and downlink data rates can be defined as,
\begin{equation}
R_m^{{ y_{\rm{1}}}} = {\log _2}\left( {1 + {P_{{y_2}}}\eta } \right){\rm{ }},m = 1, \cdots ,M,
\end{equation}
where $y_1 \in \{ \rm UL, \rm DL \}$, $ y_2 \in  \{ \rm T, \rm R \}$, $\eta  = {{{h_{m}^{{y_1}}}} \mathord{\left/{\vphantom {{{h^{{y_1}}}} {{N_0}}}} \right. \kern-\nulldelimiterspace} {{N_0}}}$. \(P_{\rm T}\) (\(P_{\rm R}\)) is the transmitting (receiving) power, and $h_m^{{y_{\rm{1}}}}$ is the channel gain between CAP and MD. When it turns to the specific \(R_{0}^{\rm UL}\) and \(R_{0}^{\rm DL}\), they are set infinitely large because computing at local CPU leaves out the process of offloading. Let \({\alpha _n}\) denote the input data size in bits, \({\gamma _n}\) is the computation data size (number of cycles required for CPU) and \({\beta _n}\) is the output data size after computation. Then, for the offloaded task $n$, the computation time, the uplink and the downlink transmission time can be
\begin{equation}
\begin{split}
t_{nm}^{\rm Comp} = \frac{{{\gamma _n}}}{{{r_m}}},{\kern 1pt}{\kern 1pt} t_{nm}^{\rm UL} = \frac{{{\alpha _n}}}{{R_m^{\rm UL}}},  {\kern 1pt}{\kern 1pt} t_{nm}^{\rm DL} = \frac{{{\beta _n}}}{{R_m^{\rm DL}}},
\end{split}
\end{equation}
where the CAP $m$ serves the tasks with a fixed rate of \({r_m}\) cycles/sec.

In fact, the CAP can start computing after either one or all scheduled tasks are offloaded. Here, we consider computation after one task offloading completes. For this case, there is no intra-CAP overlap when evaluating the overall latency. This latency is simple in expression, but the following proposed algorithm is still effective for other general expression. The three steps, offloading, computing and transmitting, take place sequentially, which results in the overall latency at CAP $m$ as follows
\begin{equation}
{T_m}\left( {\bf{X}} \right) = \sum\limits_{n \in \mathcal N} {{x_{nm}}\left( {\frac{{{\alpha _n}}}{{R_m^{{\rm{UL}}}}} + \frac{{{\beta _n}}}{{R_m^{{\rm{DL}}}}} + \frac{{{\gamma _n}}}{{{r_m}}}} \right)}.
\end{equation}
Note that since all CAPs evaluate their tasks in parallel, the delay is the maximum one, given as
\begin{equation}
T\left(\textbf X \right) = \mathop {\max }\limits_{m \in \mathcal M} {\kern 1pt} {T_m}\left(\textbf X \right).\label{overall latency}
\end{equation}

\vspace{-0.5cm}
\subsection{Energy Consumption}
An MD consumes battery to compute the tasks locally or to transmit and receive the task data. The energy consumption in the two cases can be written as
\begin{equation}
{E_1}\left(\textbf X \right) = {P_0}\sum\limits_{n \in {\mathcal {N}}} {{x_{n0}}t_{n0}^{{\rm{Comp}}}},
\end{equation}
\begin{equation}
\begin{split}
{E_2}\left(\textbf  X \right) = {\kern 1pt}&{P_{\rm T}}\sum\limits_{m \in \mathcal M\backslash \left\{ 0 \right\}} {\sum\limits_{n \in \mathcal N} {{x_{nm}}t_{nm}^{\rm UL}} } \\
&+ {\kern 1pt}{P_{\rm R}}\sum\limits_{m \in \mathcal M\backslash \left\{ 0 \right\}} {\sum\limits_{n \in \mathcal N} {{x_{nm}}t_{nm}^{\rm DL}} },
\end{split}
\end{equation}
where \({P_0}\) denotes the energy for local computation. Then, the total energy consumption is
\begin{equation}
E\left(\textbf X \right) = {E_1}(\textbf X ) + {E_2}(\textbf X).
\end{equation}

\vspace{-0.5cm}
\subsection{Optimization Problem}

Low computational latency and energy consumption are two main objectives of MEC. Unfortunately, these objectives cannot be minimized simultaneously and the problem turns out to be a multi-objective optimization. We define the weights, $\lambda_t$ and $\lambda_e$, to compromise the two objectives. Then, the weighted objective can be defined as \cite{ref1}
\begin{equation}
\Psi \left(\textbf X \right){\rm{ = }}{\lambda _t}T(\textbf X) + {\lambda _e}E(\textbf  X ),\label{Objective}
\end{equation}
where $T(\textbf X)$ is defined in (\ref{overall latency}) as the maximum delay consumed by all the CAPs instead of the sum or average one.

We aim to solve computation resource allocation scheme under specific situation where $\lambda _e$ and $\lambda _t$ are fixed. The joint minimization problem of both power and latency can be formulated as
\begin{equation}\label {jiontly function}
\begin{split}
\mathop {\min }\limits_{\bf{X}}{\kern 1pt} {\kern 1pt} & \Psi \left( {\bf{X}} \right)\\
{\rm{s.t.}}& \sum\limits_{m \in {\cal M}} {{x_{nm}}}  = 1, \forall n \in {\cal N},\\
&{x_{nm}} \in \{ 0,1\}.\\
\end{split}
\end{equation}

\vspace{-0.5cm}
\section{Offloading Learning through Cross Entropy}

The problem in (\ref {jiontly function}) is a binary integer programming one, which can be optimally solved via the branch-and-bound (BnB) algorithm with exponentially large computational complexity, especially when $\textbf X$ is large \cite{ref7}. In future wireless networks, the number of  tasks will increase and more CAPs will be involved. Then, the BnB algorithm can hardly satisfy the requirements of real-time applications. Besides, there are studies on trying to solve the problem by using conventional optimization methods. The most popular solution is to use convex relaxation, e.g. to relax \({x_{nm}} \in \left\{ {0,1} \right\}\) as \({x_{nm}} \in \left[ {0,1} \right]\) through linear programming relaxation (LPr) or to relax
\[T\left(\textbf X \right) = \mathop {\max }\limits_{m \in \mathcal M} {T_m}\left(\textbf X \right)\ \emph{\emph{as}}\  T\left(\textbf X \right) \ge \mathop {\max }\limits_{m \in \mathcal M} {T_m}\left(\textbf X \right)\]
by semidefine relaxation (SDR) \cite{ref1}. The relaxation, however, causes performance degradation compared to BnB algorithm.

Besides the above methods, the problem in (\ref{jiontly function}) with discrete optimization variables can be solved by using a probabilistic model based method, in the way of learning the probability of each policy $x_{nm}$. The CE approach is a probability learning technique in the ML area \cite{ref9}, \cite{ref10}. To solve (\ref{jiontly function}), we propose a CE approach with adaptive sampling, namely adaptive sampling cross entropy (ASCE) algorithm.

\vspace{-0.4cm}
\subsection{The CE Concept}

Cross entropy, also known in probability theory as Kullback-Leibler (K-L) divergence, serves as a metric of the distance between two probability distributions. For two distributions, $q(x)$ and $p(x)$, the CE is defined as
\begin{equation}
D( {q||p} ) = \underbrace {\sum {q( x )\ln q( x )} }_{H( q )} - \underbrace {\sum {q( x )\ln p( x )} }_{H( {q,p} )}.\label{KL}
\end{equation}
Note that in our proposed CE-based learning method $p(x)$ represents a theoretically-tractable distribution model that we try to learn for obtaining the optimal solutions, while $q(x)$ is the empirical distribution which characterizes the true distribution of the optimal solutions. Particularly, in machine learning, distribution $q(x)$ is known from observations and $H(q)$ is the entropy of $q(x)$, which leads to the equivalence of learning the CE in (\ref{KL}) and $H(q,p)$.

We are inspired by the definition of CE, a popular cost function in machine learning, to solve problem (\ref{jiontly function}) via probability learning. We learn $p(x)$ by iteratively training samples, and then generate the optimal policy of $\textbf X$ according to $p(x)$, which is close to the empirical one, $q(x)$.

\vspace{-0.5cm}
\subsection{The ASCE-based Offload Learning}

For probability learning, the probability distribution function $p(\bm x)$ is usually introduced with an indicator $\bm u$, e.g., $p(\bm x,\bm u)$ can be a Gaussian distribution and $\bm u$ contains its mean and variance \cite{ref11}. Denoting that $L$ equals to \(N \times \left( {M + 1} \right)\), the indicator $\bm u$ is a vector of $L$ dimensions, defined as \(\bm u =[u_1, u_2, \cdots , u_L] \buildrel \Delta \over =[{{\bm u}_0^{T},{\bm u}_1^{T}, \cdots ,{\bm u}_M^{T}}] \) where \({{\bm u}_m^{T}} = {\left[ {{u_{1m}},{u_{2m}}, \cdots ,{u_{Nm}}} \right]}\) and \({u_{nm}} \in \left[ {0,1} \right]\) represents the probability of $\Pr(x_{nm}=1)$. With this method, we learn $p(\bm x;\bm u)$ by learning its parameter $\bm u$. Accordingly, $\textbf X$ is vectorized as \(\bm x = \left[ {{\bm x}_0^{T},{\bm x}_1^{T}, \cdots ,{\bm x}_M^{T}} \right]\) where \({{\bm x}_m^{T}} = {\left[ {{x_{1m}},{x_{2m}}, \cdots ,{x_{Nm}}} \right]}\). Following the Bernoulli distribution, we have the distribution function $p(\bm x,\bm u)$ as \cite{ref13}
\begin{equation}
p \left( {\bm x,\bm u} \right) = \prod\limits_{l = 1}^L {{ { u_l}^{{ x_l}}}{{\left( {1 - { u_l}} \right)}^{\left( {1 - { x_l}} \right)}}}.\label{Bernoulli distribution}
\end{equation}

According to (\ref {task allocation}), one task associates to at most one CAP. Thus if a task is assigned to one CAP, its probability of being associated to other CAPs becomes zero. Aiming to reduce the redundancy of generated samples, we divide one sample, i.e., a vector $\bm x$ of $L$ dimensions, into $M+1$ independent blocks, ${\bm x}_0$, ${\bm x}_1$, $\cdots$, ${\bm x}_M$, and each of them associates to one CPU, e.g., the feasible block $[x_{1m}, \cdots, x_{Nm}]$ indicates the task assignment of tasks $1$-$N$ to CAP $m$. Let $\mathcal G$ denote the set of indices of the selected blocks in sampling and $\mathcal T$ is another set to store the samples satisfying the constraints in each iteration. We first uniformly choose $g$, an index in $\mathcal M \backslash \mathcal G$. To generate an $\bm x_g$ given $g$, we draw the entries of $\bm x_g$ according to the probability density function $p(\bm x_g, \bm u)$. For each $u_l$ in $\bm x_g$, it is drawn according to the Bernoulli distribution of parameter $u_l$. The indicator ${\bm u}_m$ of the remaining blocks in $\mathcal M \backslash \mathcal G$ is then adjusted based on ${\bm x}_g$, that is, if $x_{ig} = 1$ we have $u_{im}=0$ for $m\in \mathcal M \backslash \mathcal G$. When the cardinality of $\mathcal G$, denoted as $| \mathcal G |$, is equal to $M$, one valid sample is generated. Note that we draw the sample, while the non-feasible samples are excluded on the way. All the valid samples gather in $\mathcal T$ and the sampling repeats until the cardinality of $\mathcal T$, denoted as $|\mathcal T|$, reaches $S$.

In the proposed CE approach, computations in each iteration can be conducted in parallel, while the iterations are implemented in sequential. As will be shown later in the simulation results, we can adjust the hyperparameters of the proposed algorithm, including $S, S_{\rm elite}, \alpha$, to compromise between the amount of parallel computations per iteration and the number of iterations for convergence. This makes a flexible tradeoff between performance and latency.

\begin{table}[!t] \normalsize
\centering
\begin{tabular}{r l}
\hline
\multicolumn{2}{l}{Algorithm 1 : ASCE-based Offload Learning Algorithm}\\
\hline
1&{\textbf {Initialize:}} $\mathcal G = \mathcal T = [\ ]$, ${{\bm u}^{\left( 0 \right)}}=0.5\times{{\bm 1}_{L \times 1}}$.\\
2&{\textbf {for}} $t = 0 : T$\\
3&{\textbf {While}} $\mathcal G \ne \mathcal M $ and $ |\mathcal T| < S$\\
4&Select an index $g$ from ${\mathcal M \backslash \mathcal G}$;\\
5&Generate entries of ${\bm x}_g$ based on $p({\bm x},{\bm u})$ and update $\mathcal G$, $\mathcal T$;\\
6&Adjust ${\bm u}_m$ where $m \in ({\mathcal M \backslash \mathcal G})$ based on ${\bm x}_g$;\\
7&{\textbf {end while}} \\
8&Calculate the objective $\left\{ {\Psi \left( {{{\bm x}^s}} \right)} \right\}_{s = 1}^S $;\\
9&Sort $\left\{ {\Psi \left( {{{\bm x}^s}} \right)} \right\}_{s = 1}^S $;\\
10&Select the minimum $S_{\rm elite}$ ${\bm x}^s$ as elites;\\
11&update ${{\bm u}^{\left( t+1 \right)}}$ according to (\ref{refresh indicator in the iteration});\\
12&{\textbf {end for}}\\
13& {\textbf {Output:}} $\bm x$.\\
\hline
\end{tabular}
 \end{table}

Now we take the CE in (\ref{KL}) as the lost function. It shows that the smaller $H(q,p)$ is, the smaller the distance between $q(x)$ and $p(x)$ is. This implies
\begin{equation}\label{minimize cross entropy}
\begin{split}
\min H\left( {q,p} \right) &= \max \sum {q\left( x \right)\ln p\left( x \right)}\\
&=\max \frac{1}{S}\sum {\ln p\left( {\bm x,\bm u} \right)},
\end{split}
\end{equation}
where $q(x)$ is $\frac{1}{S}$, since the probability of each independent solution in the set of samples is $1/S$ where $S$ is the cardinality of the set \cite{ref9}. Regarding the problem in (\ref {minimize cross entropy}), the objective is equivalently to finding the optimal indicator $\bm u$ minimizing $H(q,p)$. During the $t$th iteration, $S$ series of random samples $\bm x$, serving as candidates, are drawn according to probability $p(\bm x,\bm u)$. The feasible samples generated by the adaptive sampling are under evaluation. We evaluate the objective $\left\{ {\Psi \left( {{{\bm x}^s}} \right)} \right\}_{s = 1}^S$ of (\ref{jiontly function}) and sort them as \[\Psi \left( {{{\bm x}^{\left[ 1 \right]}}} \right) \le \Psi \left( {{{\bm x}^{\left[ 2 \right]}}} \right) \le  \cdots  \le \Psi \left( {{{\bm x}^{\left[ S \right]}}} \right).\] Then, $S_{\rm{elite}}$ samples, i.e., $\bm x^{[1]}, \bm x^{[2]}, \cdots, \bm x^{[\rm elite]}$, yielding the minimum objective, are selected as elites. Now, the best indicator $\bm u$ for policy $\bm x$ can be determined as
\begin{equation}
{\bm u^*} = \arg\mathop {\max }\limits_{\bm u} \frac{1}{S}\sum\limits_{s = 1}^{S_{\rm elite}} {\ln p\left( {{\bm x^{[s]}},\bm u} \right)}.\label{fresh the indicator}
\end{equation}
Using (\ref{Bernoulli distribution}) and (\ref{fresh the indicator}) and by forcing $\frac{{\partial H(q,p)}}{{\partial {u_l}}} = 0$, the saddle point can be evaluated as
\begin{equation}
u_l^* = \frac{1}{{{S_{\rm elite}}}}\sum\limits_{s = 1}^{{S_{\rm elite}}} {x_l^{[s]}}. \label{current indicator}
\end{equation}

In the proposed learning algorithm, we choose the CE-based metric for updating the probability. Considering the randomness of sampling, especially when the number of samples is small, we update ${\bm u}^{(t+1)}$ in the $(t+1)$th iteration not only on the basis of ${\bm u}^*$ which is handled with (\ref{fresh the indicator}) and (\ref{current indicator}), but also ${{\bm u}^{\left( t\right)}}$ learned in the last iteration. It follows
\begin{equation}
{{\bm u}^{\left( {t + 1} \right)}} = \alpha {\bm u}^* + \left( {1 - \alpha } \right){{\bm u}^{\left( t \right)}},\label{refresh indicator in the iteration}
\end{equation}
where $\alpha \in \left[ {0,1} \right]$ is the learning rate  \cite{ref10}. In general, for the CE-based method, the iterations converge to an optimized solution of the problem  \cite{ref14}.

The proposed algorithm is summarized in {\textbf {Algorithm 1}}. The CE approach combining with the indicator updating mechanism can replace conventional convex optimization methods, to compromise complexity and performance.

\vspace{-0.3cm}
\section{Simulations and Discussion}

This section validates the efficiency of the proposed approach through simulations, by using the same parameters as in [1]. The MD is equipped with a CPU and $r_0=200$ Mcycles/sec, $P_0 = 0.8$ W, $P_{\rm T} = 1.258$ W and $P_{\rm R} = 1.181$ W. The CPU frequencies of the three CAPs are $r_1 = 2 \times {10^9}$, $r_2 = 2.2 \times {10^9}$ and $r_3 = 2.4 \times {10^9}$ cycles/sec. The data rates, $R_k^{UL}$ and $R_k^{DL}$, are set to be $10$ Mbps. The average objective in the figure results is the average value of the objective in (10) over a number of trials.

\begin{figure}[!t]
\centering
\includegraphics[width=2.9in]{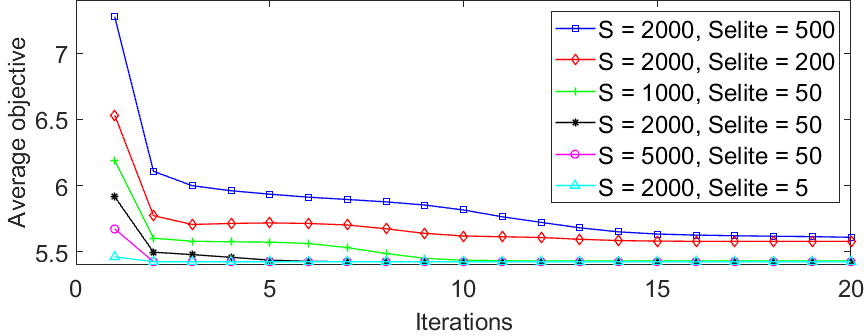}
\caption{Weighted objective against iterations with varying $S$ and $S_{\rm elite}$}
\label{fig:1}
\end{figure}

\begin{figure}[!t]
\centering
\includegraphics[width=2.9in]{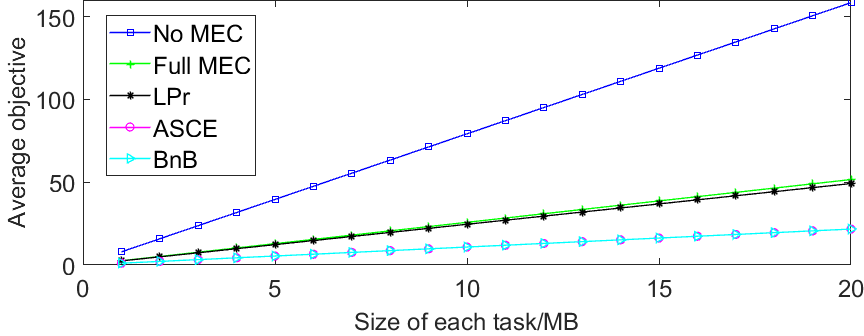}
\caption{Weighted objective under varying sizes of each task}
\label{fig:2}
\end{figure}
Fig. 1 shows the convergence of the proposed ASCE algorithm under various choices of hyperparameters $S$ and $S_{\rm elite}$. From Fig. 1, it is evident that the algorithm converges fast and the average objective reduces with $S_{\rm elite}$, which can be considered as closer to the optimum one. Moreover, the average objective converges to almost the same optimal value for all the different choices of the values of hyperparameters. We therefore conclude that, the proposed ASCE algorithm performs robustly to different values of parameters.

In Fig. 2, we compare the proposed ASCE algorithm with the LPr-based offloading algorithm in  \cite{ref1}, BnB  \cite{ref7}, No MEC and Full MEC. Among them, No MEC and Full MEC represent that all the tasks are arranged to local CPU and CAP 1, respectively. The proposed ASCE algorithm greatly outperforms the LPr method and it approaches the theoretically globally optimal solution obtained by BnB. By contrast of ``Full MEC'' and ``No MEC'', ``No MEC'' is far inferior to ``Full MEC'', which implies that the MDs of multiple tasks can work efficiently with the assist of MEC. From \cite{ref12}, the complexity of the CE approach and BnB algorithm is ${\mathcal O}\left( L \right)$ and ${\mathcal O}\left( {{2^L}} \right)$, respectively. The latter is far larger because the CE-method of parallel architecture optimizes $L$ parameters in one iteration while BnB solves parameters sequentially. Besides, the BnB algorithm requires much more memory for storage.

The offloading policy is a tradeoff between latency and energy consumption to a certain extent. The value of $\frac{{{\lambda _e}}}{{{\lambda _t}}}$ is chosen to be ${10^q}$, where $q$ grows from $-1.8$ to $2$ with step size $0.2$. While $T(\bf{X})$ plays an increasing role in the objective function, the curve presents an increasing trend for $M = 2, 3$. As for $M = 1$, there is only one CAP serving the MD, which makes the minimized latency $T(\bf{X})$ much higher than the cases with multiple CAPs. Because the minimized $E(\bf{X})$ reduces to the energy consumption of all the tasks computed locally, which is the same for all different values of $M$, the curve of $M = 1$ finally decreases to the minimized $E(\bf{X})$.

\begin{figure}[!t]
\centering
\includegraphics[width=2.9in]{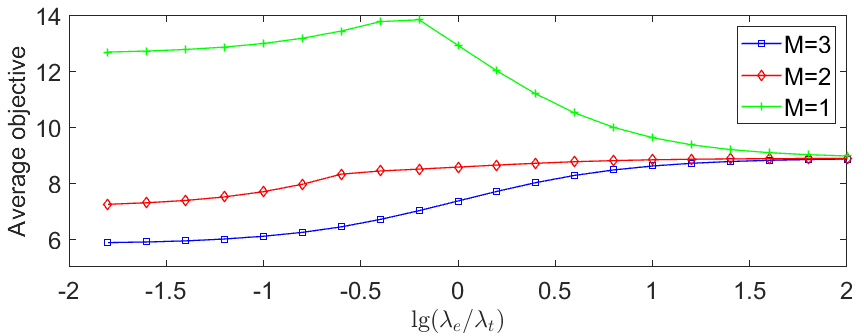}
\caption{Impacts of ratio of $\lambda_e$ to $\lambda_t$ on the weighted objective }
\label{fig:4}
\end{figure}

\vspace{-0.3cm}
\section{Conclusion}

In this paper, we present an efficient computational offloading approach for a multi-tier Het-MEC network. We propose the ASCE algorithm, which occupies less memory and has lower computational complexity than traditional algorithms. The proposed algorithm performs robustly, while it approaches closely to the optimal performance.

\end{document}